\documentclass[a4paper, 11pt]{jpconf}
\usepackage{graphicx}
\usepackage{physics}

\def\d{{\rm d}}

\def\7{$\;$}
\def\l{\left}
\def\r{\right}
\def\be{\begin{equation}}
\def\ee{\end{equation}}
\def\bea{\begin{eqnarray}}
\def\eea{\end{eqnarray}}
\def\f{\frac}

\begin{document}
\title{Towards equation of state for a market: \\ A thermodynamical paradigm of economics}

\author{Burin Gumjudpai}
\address{The Institute for Fundamental Study ``The Tah Poe Academia Institute", \\ Naresuan University, Phitsanulok 65000, Thailand}
\address{Thailand Center of Excellence in Physics, Ministry of Education, Bangkok 10400, Thailand}
\address{Graduate School of Development Economics, \\
National Institute of Development Administration (NIDA),
Bangkok 10240, Thailand}
\ead{buring@nu.ac.th}




\begin{abstract}
Foundations of equilibrium thermodynamics are the equation of state (EoS) and four postulated laws of thermodynamics. We use equilibrium thermodynamics paradigms in constructing the EoS for microeconomics system that is a market. This speculation is hoped to be first step towards whole pictures of thermodynamical paradigm of economics.
\end{abstract}
\section{Introduction}
\label{Section:Introduction}
Paradigms of thermostatics - a subject widely known as equilibrium thermodynamics, is very different from the rest of physics. It is reasonable to classify physics into two branches. The first one is the concept of equation of motion (EoM) which can be derived from least action principle. Integrations of the postulated Lagrangian over time or spacetime (in field theory) is the action varied to obtain the Euler-Lagrange equation - the EoM, of which generalized coordinates and generalized momenta are natural dynamical variables.
The other one is the concept of equation of state (EoS) in thermostatics. The EoS is constructed not theoretically like the EoM but empirically. Manifold space variables are thermodynamic coordinates such as pressure, volume, temperature, etc. without concept of time or action. We refer to the first branch as {\it dynamical laws}\footnote{This includes all quantum physics, electrodynamics and gravitational theories.} and the second one as {\it thermostatics}\footnote{This is yet to include statistical mechanics and non-equilibrium thermodynamics into this consideration.}. Considering simplest hydrostatics system, mechanical intensive coordinate ($Y$) is the minus sign of pressure ($-P$), mechanical extensive coordinate $X$ is volume ($V$), thermal intensive coordinate is temperature ($T$) and thermal extensive coordinate is entropy ($S$). Work term, $\delta W = -P\d V$ is a combination of $P$ and $V$, dubbed the {\it mechanical pair}. Heat term ($\delta \mathcal{Q} = T \d S$ is of the same spirit of the mechanical pair. Postulates of thermodynamics define $T$ (in the $0^{\text{th}}$ law), internal energy $U$ (in the $1^{\text{st}}$ law) and  $S$ (in the $2^{\text{nd}}$ law).
Constraint surface, $\Sigma$ is the EoS in form of function $f(X, Y, T)  = 0$, whereas $f(V, P, T)  = 0$ for hydrostatics system. This 4-dimensional space ($T,S,P,V$) is hence constrained into 3-dimensional space
 ($V, P, T$), ($V, P, S$), ($P, T, S$) or ($V, T, S$). Existence of the surface $\Sigma$  (the EoS), a 2-dimensional geometrical object described by 3 coordinates, enables one to define a thermodynamic potential function $U$ according to the Carath\'{e}odory's theorem (see e.g. \cite{buchdahl, munster}). It has been noticed that mathematics of economics equilibrium has lots of similarities to thermostatics\footnote{Although mathematical isomorphism between thermostatics and macroeconomics (neoclassical theory) was accepted by Samuelson but he did not accept that there are exact analogies of physical magnitudes in the economics realm \cite{sam}. A macroeconomic system inevitably relies on the concept of entropy and indeed economics system is driven by entropy (see e.g.  \cite{jk, rosser}).}. Previous investigation of this analogous
 begins with initial setup from macroeconomics \cite{mim} or with consideration of utility function as a thermodynamics potential \cite{saslow}. Here alternatively we start from the Carath\'{e}odory's axiom since, according to the axiom, the potential is to exist as the surface $\Sigma$ exists. Hence in our setup, deriving EoS of an economics system, i.e. a market, is a key foundation of our paradigm.

\section{The zeroth law of thermodynamical paradigm of economics}
 We assume there exists a 2-dimensional $\Sigma$ surface for a system which is a market, taking into account that three coordinates are quantities of demand ($Q^{\rm d}$), supply ($Q^{\rm s}$) and price ($Pr$).
  Price and temperature are both fundamentally postulated to exist before constructing theories of the subjects \cite{debrue}.
 Price is defined from market equilibrium\footnote{a condition when there is no excessive demand or excessive supply} as such the temperature is defined from thermal equilibrium. The zeroth law of thermodynamical paradigm of microeconomics should lay foundation of the concepts of market equilibrium and of price\footnote{Note that mechanism of thermal and market equilibrium formations are quite different and this is to be discussed in future work.}. The statement may read
 \begin{quote}
 {\it For three efficient, perfect competitive and clearing markets A, B, C in which there are trading of the same and homogenous normal goods. If A is in equilibrium with B and B is in equilibrium with C, hence A is in equilibrium with C. The equilibrium gives rise to existence of price function which can be ranked from lower to higher and is the character of market that determines if the market is in equilibrium with others. This type of equilibrium is the price equilibrium.}\footnote{There are other types of economics equilibrium such as interest rate equilibrium and exchange rate equilibrium.}
 \end{quote}

\section{Market with  $Q^{\rm d}$ and $Q^{\rm s}$ linear with price} \label{sec_linlin}
\begin{figure}[h]
 \includegraphics[width=18pc]{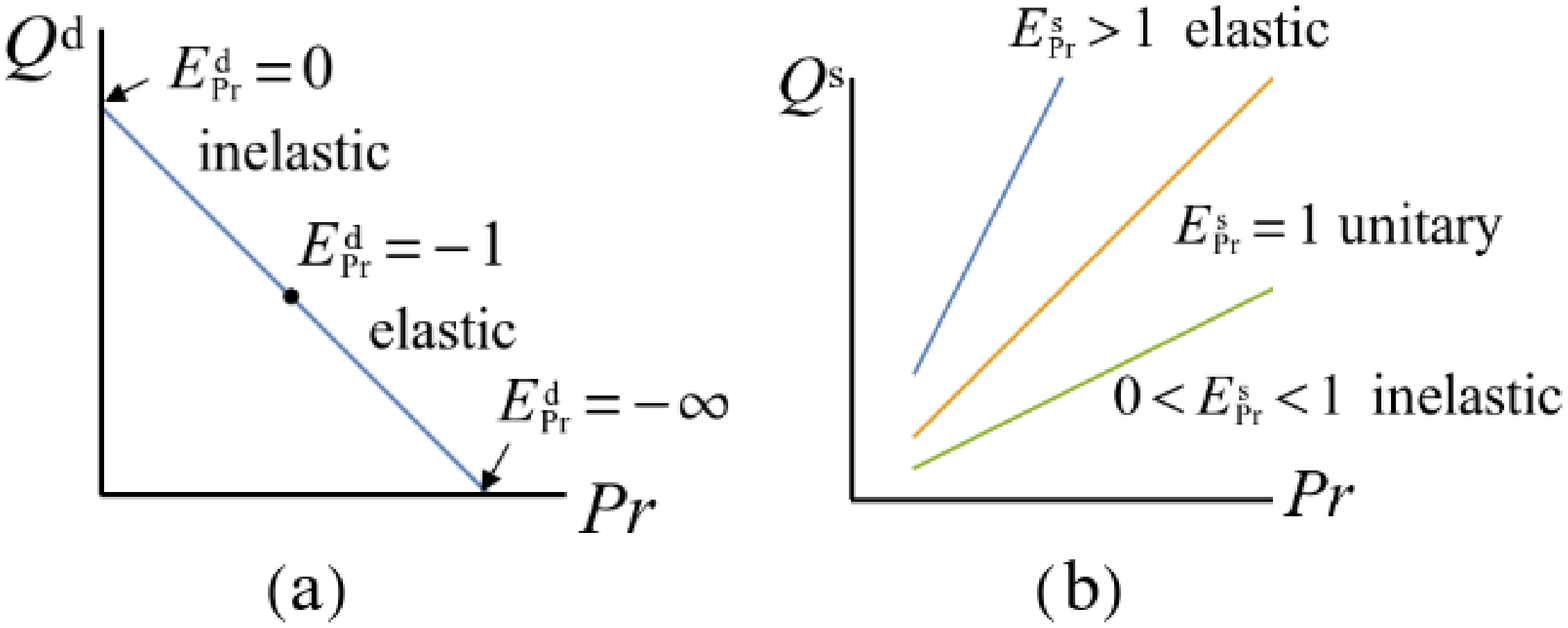}\vspace{0.1pc}\hspace{0.1pc}
 \begin{minipage}[b]{17pc}
\caption{\label{fig1} (a) linear price demand curve  (b) linear price supply curve}
\end{minipage}
\end{figure}

In a market, if the demand curve is of a function (referred to standard text books e.g. \cite{robin}),
\be
Q^{\rm d} = k_{\rm s}(Q^{\rm s})\cdot Pr + Q^{\rm d, 0} \,,   \label{eq1}
\ee
 as in figure \ref{fig1}. The proportional constant is $k_{\rm s}(Q^{\rm s}) < 0$ and the vertical intercept is $Q^{\rm d, 0}$. It could be elastic, unitary or inelastic demand depending on which section of the curve we consider. The supply curve is
$
Q^{\rm s} = k_{\rm d}(Q^{\rm d})\cdot Pr
$ as in figure \ref{fig1}.
In the same spirit with the way ideal gas law is formulated, there should be a relation
$Q^{\rm d} = k_{\rm Pr}(Pr)\cdot Q^{\rm s}$. Working with these relations, one finds $(Q^{\rm d})^2 = \varepsilon_{\rm d}^2 \cdot Pr^2  +  k_{\rm d} k_{\rm Pr} \cdot Pr\cdot Q^{\rm d, 0} $ where $\varepsilon_{\rm d}^2 = k_{\rm d} k_{\rm s} k_{\rm Pr} $ and that $(Q^{\rm s})^2 = \varepsilon_{\rm s}^2 \cdot Pr^2  +  ({k_{\rm d}}/{k_{\rm Pr}}) \cdot Pr \cdot Q^{\rm d, 0}   $ where $\varepsilon_{\rm s}^2 = \l(k_{\rm d} k_{\rm s}/k_{\rm Pr}\r) $. Point-price elasticity of demand and supply are
\be
E_{\rm d} = \f{\Delta Q^{\rm d}}{\Delta Pr}\f{Pr_0}{Q^{\rm d}_0}\,,\;\;\;\;\;\;\; E_{\rm s} = \f{\Delta Q^{\rm s}}{\Delta Pr}\f{Pr_0}{Q^{\rm s}_0}\,.   \label{eq2}
\ee
Since $k_{\rm s} < 0$ hence $\varepsilon_{\rm d}$ and $\varepsilon_{\rm s}$ are imaginary. At market clearing price, $Q^{\rm d} = Q^{\rm s}$, making $k_{\rm Pr} =1 $. The slope $\varepsilon_{\rm d} = \Delta Q^{\rm d}/\Delta Pr = E_{\rm d} Q^{\rm d}_0/Pr_0 = k_{\rm s} < 0$ in this case. The $\varepsilon_{\rm d}, \varepsilon_{\rm s}$ hence take both imaginary and real values. The inconsistency comes from existence of vertical intercept $Q^{\rm d, 0}$ in equation (\ref{eq1}) but no vertical intercept in  $
Q^{\rm s} = k_{\rm d}(Q^{\rm d})\cdot Pr
$ and from that the slope $k_{\rm d}$ and $k_{\rm s}$ are forced to be equal by market clearing condition, $Q^{\rm d} = Q^{\rm s}$.

\section{Market with unitary price demand and $Q^{\rm s}$ linear with price}
\begin{figure}[h]
 \includegraphics[width=23pc]{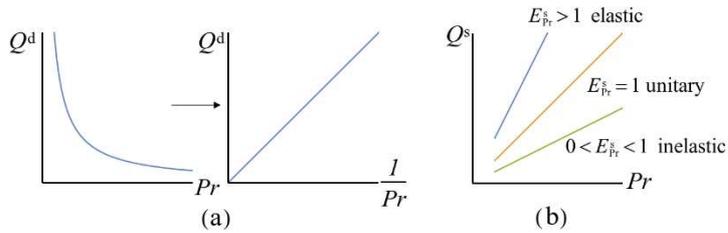}\vspace{0.1pc}
\hspace{0.1pc}
\begin{minipage}[b]{14pc}
\caption{\label{fig2}
(a) unitary price demand: curve $Q^{\rm d}$ versus $Pr$ is hyperbolic. The hyperbolic slope is $\varepsilon_{\rm d} = \varepsilon_{\rm d}(Pr)$. The curve
$Q^{\rm d}$ versus $1/Pr$ is linear. Price elasticity of demand is $E_{\rm d} = -1$ everywhere.  (b) linear price supply curve
}
\end{minipage}
\end{figure}

\begin{figure}[h]
 \includegraphics[width=12pc]{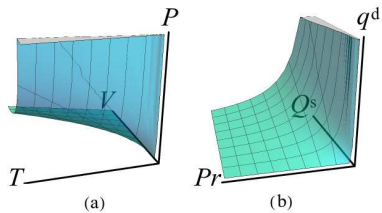}\vspace{0.1pc}
\hspace{0.1pc}
\begin{minipage}[b]{20pc}
\caption{\label{fig3} (a) EoS of ideal gas as a 2-dim. surface (b)
surface $f(Q^{\rm s}, q^{\rm d}, Pr) = 0$ (equation (\ref{eqeos})) which could be EoS of market with unitary price elastics of demand.}
\end{minipage}
\end{figure}

In distinct from Sec. \ref{sec_linlin}, the only change is the demand function,
\be
Q^{\rm d} = k_{\rm s}(Q^{\rm s})\cdot \f{1}{Pr} \,, \label{eq3}
\ee
 as in figure \ref{fig2}. Hence $k_{\rm s} > 0$ here. Note that, $\delta Q^{\rm d}  =  k_{\rm s}\cdot \delta (Pr^{-1})$ hence slope of the hyperbolic curve is $\varepsilon_{\rm d} \equiv \delta Q^{\rm d}/\delta Pr  = - k_{\rm s}/Pr^2  $. Using this slope and equation (\ref{eq3}) in equation (\ref{eq2}) resulting $E_{\rm d}  =  -1$ (i.e. unitary) at any points. Following similar procedure of Sec. \ref{sec_linlin}, we find,
$ Q^{\rm d}  = \sqrt{(k_{\rm s} k_{\rm Pr}/k_{\rm d})} Q^{\rm s}/ Pr $. Notice that the slope of supply curve is $k_{\rm d} = \varepsilon_{\rm s} = \Delta Q^{\rm s}/ \Delta Pr = E_{\rm s} Q^{\rm s}_0/Pr_{0}$.   In thermostatics, the influence (mechanical force) term is an intensive variable. We shall modify the definition of $Q^{\rm d}$ to its intensive quantity that is {\it price demand per household}, $q^{\rm d} = Q^{\rm d}/N$, where $N$ is number of household buyers. Redefining $q^{\rm d} = k_{\rm s}\cdot (1/Pr)$ and $q^{\rm d} = k_{\rm Pr} \cdot Q^{\rm s}$, we see that $k_{\rm Pr}$ is a constant $N^{-1}$. We hence succeed in writing $f(X, Y, T) = 0$ as
\be
f(Q^{\rm s}, q^{\rm d}, Pr)  = 0  \;\;\;\; \Longrightarrow\;\;\;\;  q^{\rm d}  = K \f{Q^{\rm s}}{Pr}\;\;\;\;\text{or}\;\;\;\;
Q^{\rm s}   = \f{1}{K} q^{\rm d} Pr  \label{eqeos}
\ee
where $K \equiv \sqrt{k_{\rm s}/(\varepsilon_{\rm s} N)}$ playing similar role to the gas constant $R$ or $k_{\rm B}$ combining with the specific property, e.g. compressibility factor $Z(T)$ in real virial gas. Figure \ref{fig3} presents analogous surface of ideal gas EoS and $f(Q^{\rm s}, q^{\rm d}, Pr)  = 0$ as $X \equiv Q^{\rm s}, Y \equiv q^{\rm d}$ and $T \equiv Pr$.

\section{Conclusion and critics}
This work attempts to find thermodynamical coordinates of a market. We propose the zeroth law such that the price plays the role of temperature. The EoS coordinates ($Q^{\rm s}, Q^{\rm d}, Pr$) are hypothesized. Market with linear price elasticity of demand is inconsistent due to market clearing (equilibrium) condition and form of equation. Inconsistency of $\varepsilon$ comes from the need that $Q^{\rm d} = Q^{\rm s}$ whiles having two different values of slope, $k_{\rm d}, k_{\rm s}$.
In case of market with unitary price elasticity of demand, it is possible to derive a constraint surface with an additional idea that the demand is considered to be intensive. Validity of this EoS is still our ongoing speculation.

\begin{figure}[h]
 \includegraphics[width=14pc]{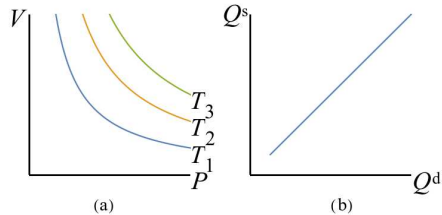}\vspace{0.1pc}\hspace{0.1pc}
\begin{minipage}[b]{22pc}
\caption{\label{fig4} (a) $T$ is an exogenous variable of the $VP$ plane. (b) In unitary demand market, $Pr$ is an exogenous however different prices do not give different curves but instead giving different value of $Q^{\rm s} = Q^{\rm d}$ via $q^{\rm d} = (1/N)Q^{\rm s}$. This is  distinct from that of hydrostatics system where $P$ and $V$ are hyperbolic to each other.}
\end{minipage}
\end{figure}

We notice that as we consider equilibrium state, quantity of goods in transaction should be denoted with only one single variable, $Q$, not to be considered separately as two forces of willing to sell and of willing to buy.
As a result, one can not draw a set of {\it isoprice curves} in the same spirit as isothermal curves (see figure \ref{fig4}). One of these variables might not be the correct coordinate. It is also noticed that thermal equilibrium is established in conservation condition. There are internal (the system) and external (the environment) parts to exchange heat and work hence resulting change in internal energy. The situation of market case is different. Transactions to form clearing price in a market
are done by buyers and sellers who are both internal agents. Forming an equilibrium price does not tell us the total wealth of the market while forming an equilibrium temperature does tell us how much internal energy the system possesses.

We also notice that in $(P, V, T)$ system,  $P$ results in change of $V$. Both are different quantities. In simple paramagnetics, magnetic field, $B_0$ induces changes in magnetization $\mathcal{M}$, via the EoS, $\mathcal{M} =  (\mathcal{D}/\mu_0) (B_0/T) $.
The physical quantities $B_0$ and $\mathcal{M}$ are similar in nature. Effect of $B_0$ is {\it amplified} by factor $\mathcal{D}$ (Curie constant) to induce $\mathcal{M}$. In unitary market, intensive influence $q^{\rm d}$ is similar to $Q^{\rm s}$ by nature of their physical quantities. From equation (\ref{eqeos}),  $(1/K)$ looks like the amplification factor for $q^{\rm d}$ to induce $Q^{\rm s}$. Here we have explored possibility towards construction of an EoS for a market. We gain some insights and learn some shortcomings of the paradigm developed here to proceed in future construction of thermodynamical paradigm of economics.

\section*{Acknowledgements}
The author thanks Y. Setthapramote (NIDA) for discussion and S. Saichaemchan for the figures. The work is supported by Naresuan University Research Grant (R2557C002).

\end{document}